# A Dual-Task Synergy-Driven Generalization Framework for Pancreatic Cancer Segmentation in CT Scans

Jun Li, Yijue Zhang, Haibo Shi, Minhong Li, Qiwei Li and Xiaohua Qian

*Abstract*—Pancreatic cancer, characterized by its notable prevalence and mortality rates, demands accurate lesion delineation for effective diagnosis and therapeutic interventions. The generalizability of extant methods is frequently compromised due to the pronounced variability in imaging and the heterogeneous characteristics of pancreatic lesions, which may mimic normal tissues and exhibit significant inter-patient variability. Thus, we propose a generalization framework that synergizes pixel-level classification and regression tasks, to accurately delineate lesions and improve model stability. This framework not only seeks to align segmentation contours with actual lesions but also uses regression to elucidate spatial relationships between diseased and normal tissues, thereby improving tumor localization and morphological characterization. Enhanced by the reciprocal transformation of task outputs, our approach integrates additional regression supervision within the segmentation context, bolstering the model's generalization ability from a dual-task perspective. Besides, dual self-supervised learning in feature spaces and output spaces augments the model's representational capability and stability across different imaging views. Experiments on 594 samples composed of three datasets with significant imaging differences demonstrate that our generalized pancreas segmentation results comparable to mainstream in-domain validation performance (Dice: 84.07%). More importantly, it successfully improves the results of the highly challenging cross-lesion generalized pancreatic cancer segmentation task by 9.51%. Thus, our model constitutes a resilient and efficient foundational technological support for pancreatic disease management and wider medical applications. The codes will be released at https://github.com/SJTUBME-QianLab/Dual-Task-Seg.

*Index Terms*—Pancreatic cancer segmentation, generalization, dual-task learning, self-supervised learning.

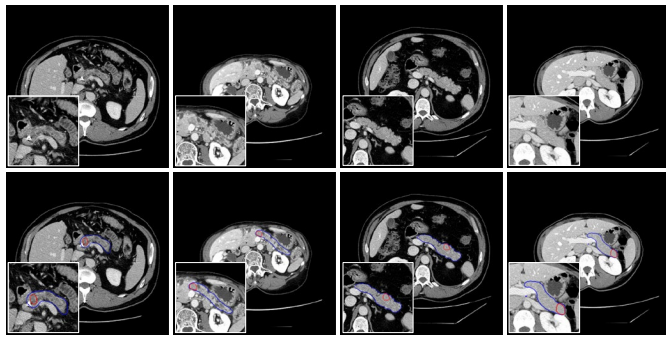

Fig. 1. Representative CT images with annotated lesions and pancreatic regions. The red and blue contours delineate the lesions and the pancreatic regions, respectively. These cases illustrate several key aspects of diagnostic imaging challenges: 1) the inherent visual disparities present in CT imaging of pancreatic lesions; 2) the often-indistinct boundaries between tumorous and normal tissues; 3) the variable presentation of tumors within the pancreas, highlighting their random occurrence and diverse locations.

## I. INTRODUCTION

PANCREATIC cancer is an aggressive malignancy frequently diagnosed at advanced stages. Surgical intervention [1], as well as advanced radiotherapy and chemotherapy [2], are significant therapeutic measures, necessitating precise segmentation of tissues. As manual segmentation becomes impractical with growing medical data, requiring extensive labor and expertise, automated lesion segmentation emerges as essential for improving diagnostic accuracy, reducing costs, and minimizing human errors [3, 4].

Nevertheless, automatic pancreatic cancer segmentation

This work was supported by the Natural Science Foundation of Shanghai under Grant 22ZR1432100, National Natural Science Foundation of China under Grant 62371286, Grant 62171273 and Grant 62401481, Fundamental Research Funds for the Central Universities under Grant 2682024CX067, China Postdoctoral Science Foundation under Grant 2024M752683. (J. Li and Y. Zhang contributed equally) (Corresponding author: M. Li, Q. Li and X. Qian).

J. Li is with the Institute of Systems Science and Technology, School of Electrical Engineering, Southwest Jiaotong University, Chengdu 611756, China, and with the Medical Image and Health Informatics Lab, School of Biomedical Engineering, Shanghai Jiao Tong University, Shanghai 200030, China (e-mail: dirk.li@outlook.com).

Y. Zhang is with the Department of Anesthesiology, Renji Hospital, Shanghai Jiaotong University School of Medicine, Shanghai，200127, China. (e-mail: zhangyijue@renji.com).

H. Shi is with the Department of Otorhinolaryngology-Head & Neck Surgery, Shanghai Sixth People's Hospital Affiliated to Shanghai Jiao Tong University School of Medicine, Shanghai, China (e-mail: haibo99@hotmail.com).

M. Li is with Department of Radiology, The Second Affiliated Hospital of Guangzhou Medical University, Guangzhou 510260, China. (minhong31@126.com)

Q. Li is with the Department of General Surgery, Renji Hospital, School of Medicine, Shanghai Jiao Tong University, Shanghai, China. (e-mail: liqiwei@renji.com).

X. Qian is with the Medical Image and Health Informatics Lab, School of Biomedical Engineering, Shanghai Jiao Tong University, Shanghai 200030, China (e-mail: xiaohua.qian@sjtu.edu.cn).

poses significant challenges due to its small size and tendency to blend with surrounding tissues. Many advanced methods [5-13] have shown promise but focus mainly on intra-domain tasks, overlooking the crucial need for generalization, which is closely linked to clinical reliability. Generalization in this context refers to a model's ability to perform well on unseen data that may vary significantly from the training data, often due to differences in patient demographics and imaging protocols across medical institutions. In fact, traditional machine learning methods typically assume data is independent and identically distributed (IID) [14], expecting uniformity across training and test datasets. However, this assumption renders models sensitive to data variability (Fig. 1), a significant issue in clinical settings where data originates from varied medical institutions with distinct patient demographics and imaging protocols. Such variability challenges the IID assumption, compromising model accuracy and stability. In this work, we aim to enhance model performance on unseen datasets with diverse appearances, addressing this critical limitation.

However, the demand for improved generalization in pancreatic cancer segmentation amplifies the task's complexity, especially given its already challenging nature. Consequently, the primary limitation of existing methods lies in their insufficient generalizability to support stable performance of models across multiple datasets. In this work, we attribute these generalization challenges primarily to differences in anatomical features and imaging characteristics. Lesions within the pancreas display significant variability in location, shape, and texture across diverse imaging contexts, disease stages, and patient cohorts. We distill these challenges into three key areas:

1) **Significant appearance discrepancies.** Remarkable differences in pancreatic texture and appearance (Fig. 1) arise from demographic, technological and procedural variations across medical centers, highlighting discernible differences even within the same dataset [15]. Thus, this significant appearance variability poses a substantial challenge to the model's generalization stability.

2) **Unclear tumor contours.** Contour irregularities are exacerbated by the aggressive progression of pancreatic cancer, which frequently infiltrates into surrounding normal tissues, creating ambiguous boundaries (Fig. 1). Besides, the small size further complicates accurate delineation, with minor contour deviations significantly impacting accuracy. Moreover, models often rely on anatomical shapes and grayscale texture to trace boundaries, but the variability of these features intensifies the difficulty in producing precise contours on unseen data, consequently impairing potential generalization stability.

3) **Random anatomical locations.** The random locations of lesions present a significant challenge for models that use fixed relative positions of abdominal organs as anatomical priors [16]. The variability in lesion occurrence within the pancreas, as illustrated in Fig. 1, undermines the utility of contextual anatomical features in developing these models. Consequently, the absence of consistent anatomical markers may result in models that are overly adapted to specific appearance patterns, thereby substantially impairing their generalization capabilities across datasets with diverse appearance variations.

The three primary challenges could be mitigated as follows:

a) **Dealing with significant appearance discrepancies:** While data augmentation techniques [17] have shown capability in addressing data variability, their limited scope in introducing texture variations restricts their effectiveness in improving generalization for tumor segmentation. Thus, a more effective appearance perturbation method is required, one that can introduce significant texture variations without compromising the integrity of anatomical structures.

b) **Dealing with unclear tumor contours:** The texture perturbation approach lacks the capacity to enrich boundary delineation information, leaving the precise tracing of tumor edges as a persistent challenge. Although direct boundary prediction has proven successful [18], its applicability to pancreatic lesions is hindered by their diminutive size and insufficient anatomical detail. Thus, an attempt can be made to incorporate more supervisory signals to augment boundary error correction, elevating the quality of segmentation contours.

c) **Dealing with random anatomical locations:** The unpredictable locations of lesions necessitate a reliance on texture rather than unstable anatomical cues for lesion identification. Given the significant variation in texture features across datasets, which often results in misidentification, efforts can be made to harness additional spatial information to rectify inaccuracies in tumor location prediction, while simultaneously bolstering the model's adaptability to texture variations.

Thus, we propose a dual-task synergy-driven generalization framework for pancreas cancer segmentation, designed to enhance model robustness and segmentation accuracy on novel data characterized by diverse appearances, tumor locations, and lesion boundaries. This framework also integrates dual self-supervised learning strategies across feature and output spaces to stabilize tumor segmentation amidst extensive data variability. Specifically, by applying random convolution operators, we generate sample pairs with identical anatomical structures yet varied grayscale textures, fostering model adaptability to unseen grayscale textures. Besides, binary masks are transformed into distance maps, which delineates the distance of each pixel from the tumor boundary, conveys pixel class information, and provide spatial neighborhood context. Indeed, these distance maps could enrich spatial context, aiding in precise tumor localization and boundary fitting. Thus, they are used to supervise the pixel-level regression task in collaboration with the segmentation task. To synchronize segmentation and regression efforts for improved stability, we facilitate the reciprocal conversion of task outputs, ensuring regression adjustments are informed by segmentation insights. Finally, cross-instance contrastive learning and consistency learning are established in both the feature space and the output space, promoting uniform feature representations and consistent, reliable outcomes across variable unseen images. The main contributions of this work are as follows:

- **A generalization framework**. It addresses three task-specific confounding factors, exhibiting greater potential for clinical applications.

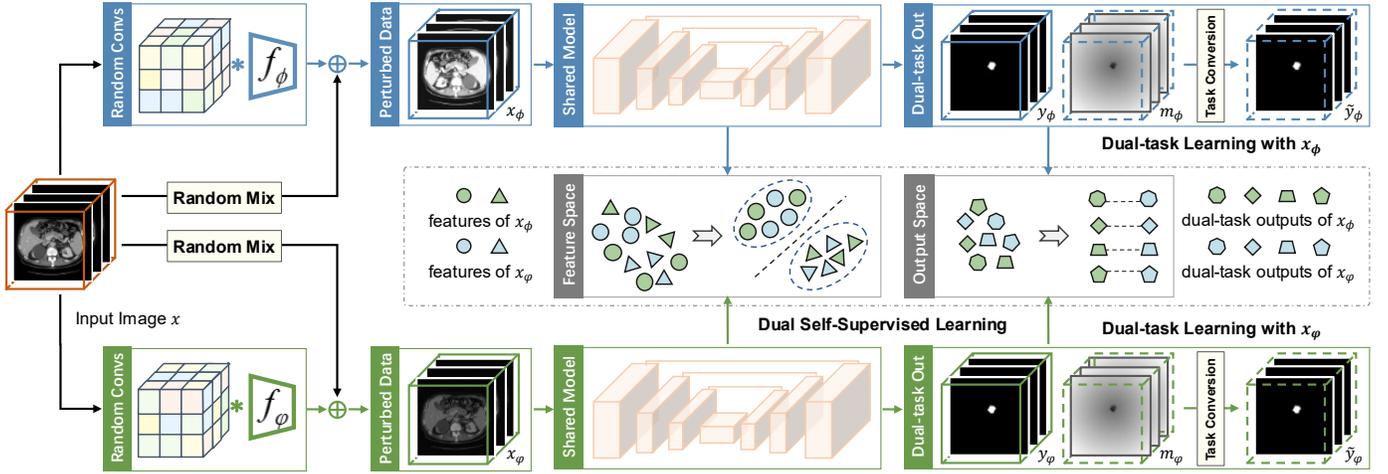

Fig. 2. Schematic of the dual-task synergy-driven generalization framework. Initially, the images undergo a random augmentation process driven by two stochastic convolution operators, producing pairs of images that are structurally identical but exhibit substantial variations in appearance and texture. They are then concurrently fed into the segmentation network, where consistency optimization is carried out both in the feature space and the output space. Finally, the segmentation and regression outcomes are further refined through a task transformation module, facilitating cross-task learning. This approach aims to leverage insights from multiple space and tasks to significantly improve the model's generalization capabilities across diverse imaging scenarios.

- **A dual-task synergy-driven strategy**. It incorporates rich spatial information into pixel-level tasks of classification and regression. The amalgamation of the two task perspectives bolsters segmentation stability for various data distributions.
- **A dual self-supervised learning module**. It incorporates contrastive learning and consistency learning in the feature and output spaces. This ensures consistent characterization and stable segmentation under wide variations in unseen data.

## II. RELATED WORK

### A. Pancreas and Pancreatic Cancer Segmentation

The pancreas and its lesions are widely acknowledged as among the most challenging regions within abdominal tissue segmentation tasks [19]. Traditional methods for pancreas segmentation [20-22] employed a strategy of aggregating information through the fusion of results from coronal, sagittal, and axial planes, leveraging 2D models to approximate a three-dimensional understanding. However, with advancements in computational capabilities and algorithmic methodologies, recent research endeavors [23, 24] have shifted towards directly capturing spatial information through the application of 3D networks. This paradigm shift enables a more nuanced and accurate representation of the complex anatomical structures of the pancreas and its lesions. For the more challenging task of pancreatic cancer segmentation, early research predominantly focused on developing automatic algorithms within specified regions of interest (ROI), such as Gaussian mixture models [25], multi-channel models [5], local texture enhancement models [6]. To avoid manually outlining ROI, modern methods are directed towards developing end-to-end models. These advancements encompass coarse-to-fine strategies [7, 26], designed to incrementally improve segmentation precision; neural architecture search methods [8, 9], aimed at automating the creation of networks specifically optimized for segmentation; and the employment of multimodal data [5, 11, 27], which has been leveraged to significantly bolster the accuracy of tumor segmentation. While these methods have marked substantial advancements, their validation has predominantly occurred on localized datasets, presenting obstacles in affirming their stability and reliability for practical applications. Consequently, our objective is to develop a generalization framework that necessitates only a single training phase yet ensures consistent performance, thereby rendering it more feasible and accessible for deployment in real-world settings.

### B. Generalization Methods for Medical Images

Generalization methods aim to help models to perform reliably on unseen data, which reflects the challenges of real-world scenarios. To enhance the generalization performance, some methods aim to generate additional samples with diverse appearance patterns, such as deep stacked transformations [17], frequency-domain space perturbation [28], and random linear combination of samples [29]. Delisle et al. [30] followed a different strategy, where a universal normalization function was constructed to alleviate the impact of different imaging styles. Moreover, model-independent training paradigms, such as the model-agnostic meta-learning strategies [31, 32], invariant-content collaborative learning strategy [33], prior knowledge pool-based feature enhancement strategy [34], domain- and content-adaptive convolution [35], and domain fusion-based attention mechanism [36], were proposed to improve the generalization performance. However, most of these methods were designed for organ segmentation and didn't consider the challenges of lesion segmentation. Thus, we aim to develop strategies tailored to construct a pancreatic cancer segmentation model with improved generalization performance.

## III. METHODS

### A. Texture Enhancement

Neural networks typically rely on local texture and other surface features for accurate predictions [37]. Such networks typically perform well when the feature differences between the training and test sample distributions can be ignored. However, this assumption can be problematic in medical imaging, where texture variations due to different patient populations and imaging factors can affect generalization stability. In this work, this issue is alleviated through appearance perturbations to broaden the range of the image variations and improve the model adaptability to wide data variability in clinical scenarios.

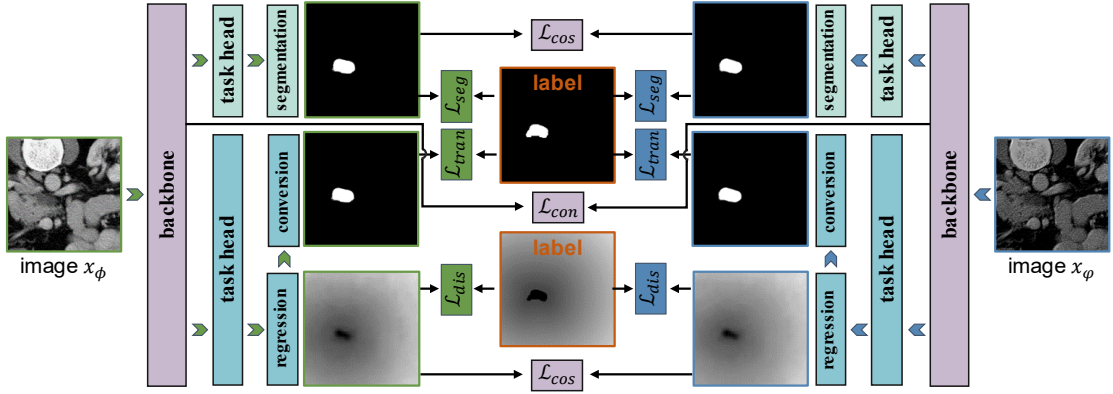

Fig. 3. An overview of the training process for the proposed multi-head model for joint segmentation and regression. Images $x_\phi$ and $x_\varphi$, which have undergone random enhancements to possess identical structures but different appearances, are input into a backbone network. This network outputs corresponding segmentation and regression results through two distinct task branches. Segmentation loss and regression loss are denoted by $\mathcal{L}_{seg}$ and $\mathcal{L}_{dis}$, respectively. Consistency optimization within the feature space and output space is driven by $\mathcal{L}_{con}$ and $\mathcal{L}_{cos}$, respectively. Additionally, cross-task learning loss, represented by $\mathcal{L}_{tran}$, integrates insights across both tasks to enhance the model's overall performance and accuracy in diverse imaging conditions.

Convolution operations can perturb local texture without altering the main structure [38, 39], thus a single convolutional operator can be used to enrich image texture. As shown in Fig. 2, each sample is separately acted upon by two convolutional operators to produce different grayscale texture patterns. Each operator is realized as a random-parameter-driven convolution. Specifically, let the image be $x$, and let two convolutional operators be denoted by $F_\phi$ and $F_\varphi$, respectively. A pair of data samples with the same anatomy but with different textures can be obtained through applying this pair of operators. To preserve the original anatomical structure, convolved image is randomly mixed with the original image $x$ to obtain the sample pair $(x_\phi, x_\varphi)$. This process can be expressed as follows:

$$x_\phi = a_\phi \cdot F_\phi(x) + (1 - a_\phi) \cdot x, \quad a_\phi \sim U(0,1),$$
$$x_\varphi = a_\varphi \cdot F_\varphi(x) + (1 - a_\varphi) \cdot x, \quad a_\varphi \sim U(0,1), \quad (1)$$

### B. Dual-task Learning
#### 1) Segmentation task
The segmentation loss is defined as the sum of the Dice loss and the binary cross-entropy loss:

$$L = 1 - \frac{2 \times \sum_{i=1}^{N} Y_i \hat{Y}_i}{\sum_{i=1}^{N} Y_i + \sum_{i=1}^{N} \hat{Y}_i} - \frac{1}{N} \sum_{i=1}^{N} \left[ Y_i \log(\hat{Y}_i) + (1 - Y_i) \log(1 - \hat{Y}_i) \right], \quad (2)$$

where $i$ denotes the pixel index, $N$ is the total number of pixels, $Y$ and $\hat{Y}$ represent the ground-truth and segmentation maps, respectively. Since the network input is structured as ample pairs $(x_\phi, x_\varphi)$, each result contains two parts: $y_\phi$ and $y_\varphi$. To minimize omissions, a slice-level classification loss was incorporated to determine whether each slice contains tumor tissue. Let $c_x$ denote the prediction and $C$ denote the label, the segmentation loss $\mathcal{L}_{seg}$ can be expressed as follows:

$$L_{seg} = L(y_\phi, Y) + L(y_\varphi, Y) + L(c_\phi, C) + L(c_\varphi, C), \quad (3)$$

#### 2) Regression task
Pancreatic cancer lesions typically exhibit invasive behavior, infiltrating adjacent healthy tissues, resulting in indistinct boundaries between diseased and normal tissues. Consequently, segmentation models often produce blurred or noisy results, particularly at the edges. This imprecision has a more pronounced effect on small-volume pancreatic cancer lesions, where minor contour variations can significantly degrade segmentation metrics. Hence, there is a need for more robust segmentation methods to produce precise tumor contours.

To this end, we propose converting binary masks into distance maps, providing a continuous representation of the spatial neighborhood relationships between all pixels and the tumor boundary. It characterizes the voxel's class, provides more spatial information, and constrains the segmentation process to generate more reasonable boundaries. The distance map construction depends on a signed distance function that assigns negative and positive values to voxels inside and outside the target region, respectively. Each value indicates the shortest distance from a voxel to the target boundary. The distance map elements can be expressed as follows:

$$M_i = \begin{cases} -E(Y_i, \partial Y), & Y_i \in Y_{in}, \\ +E(Y_i, \partial Y), & Y_i \in Y_{out}, \end{cases} \quad (4)$$

where $\partial Y$ represents the boundary of $Y$, $Y_{in}$ and $Y_{out}$ represent voxels inside and outside the lesion, and $E(\cdot)$ represents the Euclidean distance. We then introduce an integrated approach that combines distance regression with segmentation utilizing the distance map $M$. It employs a multi-head network (Fig. 3), featuring a convolutional block integrated into the output layer. The block enforces a constraint to promote consistency between the predicted and ground-truth distance maps. This enhances the precision of spatial information used in segmentation and promotes synergy between the segmentation and distance regression tasks. The regression task is guided by the $L_1$ loss between the predicted and generated distance maps:

$$L_{dis} = \frac{1}{2|x|} \sum \left[ \left| D(F_\phi(x)) - M \right| + \left| D(F_\varphi(x)) - M \right| \right], \quad (5)$$

where $D(\cdot)$ denotes the distance regression network.

### 3) Task conversion

As the distance map is derived from the binary segmentation mask, it implies that the distance regression results can be translated into segmentation results. Given the complementary relationship between regression and segmentation, there is potential for mutual enhancement, highlighting the significance of their convertibility. To facilitate cross-task joint optimization, we transform the distance maps $(m_\phi, m_\varphi)$ into segmentation maps $(\tilde{y}_\phi, \tilde{y}_\varphi)$. It provides guidance on the regression using segmentation labels, leading to more efficient joint optimization. The transformation can be expressed as follows:

$$\tilde{y}_\phi^i = \begin{cases} 1, & if \ m_\phi^i \leq 0, \\ 0, & otherwise, \end{cases} \quad (6)$$

where $i$ denotes the $i^{th}$ voxel. The transformed segmentation outcomes $\tilde{y}_\phi$ and $\tilde{y}_\varphi$ can be used to guide progress on the distance regression task with segmentation labels. Such cross-task supervision can be expressed with the following loss:

$$L_{tran} = \frac{1}{2} \left[ L_{seg}(\tilde{y}_\phi, y) + L_{seg}(\tilde{y}_\varphi, y) \right], \quad (7)$$

### C. Dual Self-supervised Learning

We first apply a contrastive learning strategy in the feature space. Samples are obtained from the feature maps of $x_\phi$ and $x_\varphi$. The feature maps are derived from the encoder's output and whose dimension is reduced by a projection layer. Owing to the spatial consistency between the feature map and the label, the positive samples $f_p$ and negative samples $f_n$ can be obtained from specific locations in the feature maps using labels, i.e., $f_p$ is located at the central tumor regions, while $f_n$ is derived from normal tissues surrounding the tumors. This not only helps the model to distinguish between tumorous and normal tissues in the same sample, but also enables cross-instance recognition, leading to consistent feature representations in $x_\phi$ and $x_\varphi$ with different grayscale texture patterns. Assuming that the number of positive and negative samples in each minibatch is $B$, the total number of samples obtained from $x_\phi$ and $x_\varphi$ is $2 \times 2 \times B$. The cross-instance contrastive learning loss can be expressed as:

$$L_{con} = \sum_{i=1}^{4B} \sum_{j=i+1}^{4B} \frac{1}{2 \times N_C} \left( 1 - \mathbb{F}(f_i, f_j) \right) L_{inf}(f_i, f_j), \quad (8)$$

where $\mathbb{F}(f_i, f_j)$ is an indicator function whose value is 0 when $f_i, f_j$ are of the same class, but 1 otherwise. $N_C$ is the overall count of the potential pairs and varies across samples due to differences in tumor sizes. $\mathcal{L}_{inf}$ is the InfoNCE loss [40]:

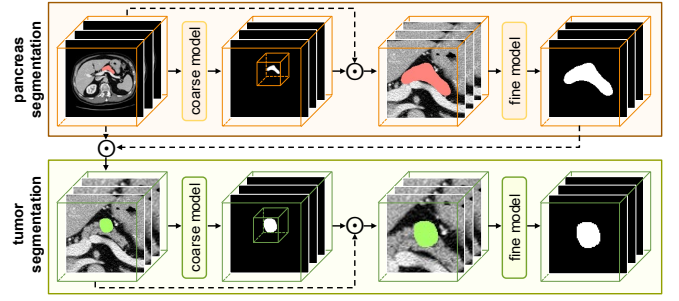

Fig. 4. Workflow for generalized segmentation of the pancreas and pancreatic cancer. Both pancreas segmentation and pancreatic cancer segmentation adhere to a coarse-to-fine segmentation process. ROIs are first identified through coarse segmentation, and then undergo fine segmentation to yield precise anatomical details. Furthermore, the results of the pancreas segmentation are utilized as the ROI for the coarse segmentation of pancreatic cancer. This structured approach ensures a detailed and accurate delineation of both the pancreas and pancreatic cancer.

$$L_{inf}(f_i, f_j) = -\log \frac{e^{L_{sim}(f_i, f_j)/\tau}}{\sum_{i=1, j\neq i}^{4B} \mathbb{F}(f_i, f_j) e^{L_{sim}(f_i, f_j)/\tau}}, \quad (9)$$

where $\tau$ denotes the temperature scaling parameter and $\mathcal{L}_{sim}(a,b) = a^T b / \|a\| \|b\|$ measures the cosine similarity.

Self-supervised learning in the output space is guided by a consistency constraint. Since $x_\phi$ and $x_\varphi$ differ only in the grayscale texture but have the same anatomical structure, their segmentation results should be consistent. As this consistency gets higher, the model becomes more stable and less affected by grayscale texture variations. In addition, for the distance regression task, consistent regression results are sought for different appearance styles. Thus, a cross-instance consistency loss is employed for the segmentation and regression tasks:

$$L_{cos} = \frac{1}{|x|} \left( \sum \left\| \hat{y}_\phi - \hat{y}_\varphi \right\|^2 + \sum \left\| m_\phi - m_\varphi \right\|^2 \right), \quad (10)$$

Let $L_{ssl}$ be the sum of $\mathcal{L}_{con}$ and $\mathcal{L}_{cos}$, the overall loss consists of four main components (segmentation, regression, cross-task supervision, and cross-instance self-supervised learning):

$$L_{all} = L_{seg} + L_{dis} + L_{tran} + L_{ssl}, \quad (11)$$

### D. Learning Strategies and Network Architecture

Fig. 4 depicts the workflow of our method, which has two distinct stages, each adopting a coarse-to-fine approach. Initially, the pancreas is subjected to coarse and then fine segmentation processes. Next, the pancreas segmentation outcomes are used to identify each external bounding box as a pancreatic ROI. Given the substantial size disparity between the pancreas and lesions, tumor segmentation presents considerable challenges and necessitates further refinement. Hence, we also establish a fine model that performs secondary segmentation in the ROI created based on the initial tumor segmentation. The testing strategy of coarse segmentation and fine segmentation are consistent with previous studies [23]. As illustrated in Fig. 2, a 3D encoder-decoder structure [41] serves as the shared model. This backbone comprises four residual convolutional layers [42]. The decoder pathway employs trilinear interpolation to avoid checkerboard artifacts potentially introduced by transposed convolutions. A dropout operator with a probability of 0.1 is incorporated within the bottleneck layer. Task-specific heads, implemented with a convolutional layer, compress the 32-channel feature maps from the bottleneck into a single-dimensional output.

## IV. RESULTS

### A. Dataset Description

We employed three datasets for evaluation: the Medical Segmentation Decathlon (MSD) dataset [15], the RMYY dataset from Jiangsu Province Hospital, and the RENJI dataset from Renji Hospital. This study was approved by Renji Hospital Ethics Committee (IRB No. RA-2021-094). Table I summarizes key information regarding these datasets, including details about patient cohorts, imaging protocols, and scanner specifications. The MSD dataset includes cases involving pancreatic ductal adenocarcinoma (PDA), intraductal papillary mucinous neoplasms (IPMN), and pancreatic neuroendocrine tumors (PNET). Given the large variation in slice thicknesses, resampling was applied to achieve a uniform spatial resolution of 1.5×0.8×0.8 to mitigate the effects of anisotropy on the 3D network. The RMYY and RENJI datasets predominantly comprise PDA cases. All tumor labels were manually annotated, while pancreatic labels in RMYY dataset were initially generated automatically, then manually refined, and confirmed by experienced physicians. Imaging equipment varied across these centers, with tube voltage settings ranged from 110 to 140 kVp, and tube current spanned from 100 to 545 mAs across datasets, reflecting the diverse imaging conditions. Spiral pitch factors also varied, ranging from 0.6 to 1.375, which introduces additional variability in the imaging data. To enhance the generalizability of our pancreas segmentation model, we additionally incorporated the NIH dataset [19], which contains normal pancreas cases without lesions. Across all datasets, intensity values were truncated to the range of [-100, 240] Hounsfield units and normalized for consistency.

TABLE I
KEY INFORMATION OF PANCREATIC CT IMAGING DATASETS.

| Parameters | MSD | RENJI | RMYY |
|---|---|---|---|
| Cases | 281 | 160 | 153 |
| Patient Cohort | PDA/IPMN/PNET | PDA | PDA |
| Slices | 37-751 | 152-684 | 41-336 |
| Spacing (mm) | 0.60-0.97 | 0.55-0.88 | 0.54-0.86 |
| Thickness (mm) | 0.70-7.50 | 0.59-1.25 | 1.50-5.00 |
| Manufacturer | GE | SIEMENS/GE/ TOSHIBA/ Philips | GE/Philips/ SIEMENS |
| Tube Voltage (kVp) | 120 | 110/120/130 | 110/120/140 |
| Tube Current (mAs) | 220-380 | 102-380 | 100-545 |
| Spiral Pitch Factor | 0.984-1.375 | 0.6-1.375 | 0.6-1 |

### B. Evaluation Metrics

We evaluate segmentation performance using the Dice Similarity Coefficient (DSC), the Average Symmetric Surface Distance (ASD), and the 95th-percentile Hausdorff Distance (HD). DSC is often reported as a percentage and provides a measure of volumetric overlap. In contrast, ASD and HD reflect boundary-level discrepancies and are reported in millimeters. While ASD captures the average boundary deviation between predicted and ground-truth surfaces, HD is more sensitive to outliers, highlighting the maximum local boundary mismatch. It should be noted that ASD and HD may not be computable for certain cases with extremely small lesions. Under such circumstances, we assign standardized values of 40 mm for ASD and 100 mm for HD.

### C. Implementation Details

All methods were implemented using PyTorch on two NVIDIA A30 GPUs. An Adam optimizer was used with a batch size of 1 and an initial learning rate of $1 \times 10^{-4}$, which exponentially decayed as $lr_{i+1} = lr_i \times ((1 - current\_iter/max\_iter)^{0.9})$. Given the computational expense of pancreas segmentation and its function as ROI delineation, the coarse and fine models incorporate only selected modules and are trained for a maximum of 150 epochs to enhance efficiency. The full model was reserved for tumor segmentation, with the coarse and fine models trained for a maximum of 800 epochs. Data augmentation included random rotations (−15° to 15°), mirroring, and noise addition. The generalization experiments are conducted using a round-robin strategy wherein two of the three datasets were always used for training, and the third dataset was reserved as an unseen external test set. This setup was meticulously rotated to ensure that each dataset was used as the test set in turn, thus allowing us to thoroughly assess the model's ability to generalize across different domain shifts represented by varying data centers.

TABLE II
GENERALIZED PANCREAS SEGMENTATION RESULTS WITH A COARSE-TO-FINE STRATEGY.

| Stage | RMYY | | | MSD | | | RENJI | | | Mean | | |
|---|---|---|---|---|---|---|---|---|---|---|---|---|
| | DSC (%) | ASD (mm) | HD (mm) | DSC (%) | ASD (mm) | HD (mm) | DSC (%) | ASD (mm) | HD (mm) | DSC (%) | ASD (mm) | HD (mm) |
| Coarse | 78.74 | 2.60 | 8.53 | 77.55 | 2.64 | 9.65 | 78.32 | 2.25 | 8.19 | 78.20 | 2.49 | 8.79 |
| Fine | 85.99 | 1.78 | 6.66 | 82.23 | 2.06 | 7.98 | 84.01 | 1.66 | 6.21 | 84.07 | 1.83 | 6.95 |

### D. Generalized Pancreas Segmentation

The method outlined employs a sequential workflow that initiates with pancreas segmentation before proceeding to tumor segmentation (Fig. 4). Thus, the precision of pancreas segmentation is paramount for establishing a dependable pancreatic ROI; inaccuracies in this phase could detrimentally impact the subsequent tumor segmentation outcomes. In this context, Table II showcases the generalized pancreas segmentation performance, where the terms 'Coarse' and 'Fine' correspond to the initial and advanced stages of segmentation, respectively. Notably, the 'Fine' stage demonstrates superior performance, thereby mitigating the risk of inaccurate tumor segmentation due to flawed ROIs. According to two recent review studies [43, 44] and a large-scale multi-center study [45], the state-of-the-art pancreas segmentation performance—with large-scale data and internal cross-validation settings—stands at 88.31%. By contrast, results on external, independent validation datasets vary from 67% to 83.7%. These findings confirm that our generalized pancreas segmentation results surpass the performance of all models evaluated on external test sets. Notably, these studies employed significantly larger training cohorts than ours, thereby strongly supporting both the superiority and the robustness of our method.

### E. Generalized Pancreatic Cancer Segmentation

We compared our model with the intra-domain cross-validation model, which shares the same backbone network and represents the upper limit of generalized model performance. Besides, tumor segmentation can be affected by the accuracy of pancreatic ROI. To assess tumor segmentation independently based on model capabilities, we used both pancreas segmentation outcomes and pancreas labels as pancreatic ROIs. We use 'Seg' and 'Mask' to denote the methods of obtaining pancreatic ROIs from segmentation results and ground-truth labels, respectively. Moreover, we denote coarse and fine segmentation as 'C-' and 'F-', respectively. We end up with four distinct configurations to consider, namely: 'C-Seg', 'F-Seg', 'C-Mask' and 'F-Mask', wherein each is characterized by its ROI method and segmentation stage.

TABLE III
GENERALIZED PANCREATIC CANCER SEGMENTATION RESULTS BASED ON ROIS OBTAINED FROM PANCREAS SEGMENTATION AND MANUAL MASK, RESPECTIVELY.

| Method | RMYY | | | MSD | | | RENJI | | | Mean | | |
|---|---|---|---|---|---|---|---|---|---|---|---|---|
| | DSC (%) | ASD (mm) | HD (mm) | DSC (%) | ASD (mm) | HD (mm) | DSC (%) | ASD (mm) | HD (mm) | DSC (%) | ASD (mm) | HD (mm) |
| C-Seg | 42.78 | 9.50 | 21.01 | 31.64 | 18.49 | 33.20 | 47.67 | 9.77 | 23.08 | 40.69 | 12.58 | 25.76 |
| F-Seg | 48.36 | 8.62 | 17.67 | 36.85 | 18.06 | 28.60 | 53.12 | 8.98 | 22.10 | 46.11 | 11.88 | 22.79 |
| C-Mask | 44.76 | 8.83 | 19.72 | 34.86 | 15.93 | 28.30 | 46.95 | 10.19 | 24.00 | 42.73 | 11.64 | 24.00 |
| F-Mask | 50.06 | 7.96 | 16.24 | 38.75 | 15.23 | 24.84 | 53.18 | 9.25 | 22.80 | 47.33 | 10.81 | 21.29 |

TABLE IV
PANCREATIC CANCER SEGMENTATION RESULTS WITH FOUR-FOLD CROSS-VALIDATION STRATEGY AND NO DOMAIN SHIFT.

| Method | RMYY | | | MSD | | | RENJI | | | Mean | | |
|---|---|---|---|---|---|---|---|---|---|---|---|---|
| | DSC (%) | ASD (mm) | HD (mm) | DSC (%) | ASD (mm) | HD (mm) | DSC (%) | ASD (mm) | HD (mm) | DSC (%) | ASD (mm) | HD (mm) |
| C-Seg | 46.67 | 9.98 | 19.00 | 44.89 | 13.75 | 24.12 | 55.78 | 9.27 | 22.48 | 49.11 | 11.00 | 21.86 |
| F-Seg | 48.22 | 10.57 | 21.01 | 43.86 | 14.66 | 24.72 | 57.67 | 9.27 | 22.93 | 49.91 | 11.50 | 22.88 |
| C-Mask | 47.13 | 9.42 | 18.94 | 46.42 | 13.15 | 23.61 | 56.70 | 9.88 | 22.54 | 50.08 | 10.81 | 21.69 |
| F-Mask | 48.74 | 10.31 | 20.46 | 44.48 | 14.17 | 24.19 | 58.23 | 9.75 | 22.64 | 50.48 | 11.41 | 22.43 |

Table III and Table IV present the results of the four cases above for the generalization and intra-domain models. The acceptable gap indicates that our approach effectively enhances the generalization performance to closely match the performance of the intra-domain cross-validation model. Fig. 5 shows visual segmentation results. The fourth column shows accurate pancreas segmentation by our model, yielding reliable pancreatic ROIs. The fifth and sixth columns depict coarse and fine segmentation outcomes, demonstrating the efficacy of our two-stage approach. Furthermore, even in situations where coarse segmentation fails completely (the second row), the fine-segmentation model can compensate for this failure and achieve outstanding performance. In addition, Table III shows that our model performs comparably with segmentation-based and label-based ROI, with only a 1.22% difference in average DSC. This demonstrates our model's adaptability to varying pancreatic ROI qualities and highlights the similarity between the segmentation results and the labels. Thus, our method shows good generalization performance for the segmentation of pancreas and the pancreatic lesions.

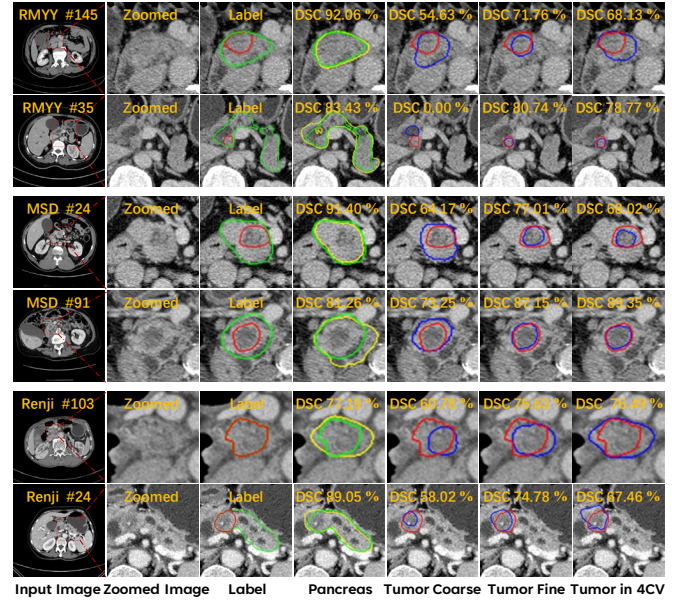

Fig. 5. Visual samples. The red and green contours denote the labels for tumor and pancreas, respectively. The blue and yellow contours denote the segmentation results for tumor and pancreas, respectively. "Tumor Coarse" and "Tumor Fine" denote the generalized coarse and fine segmentation of lesions, respectively. "Tumor in 4CV" denotes the four-fold cross-validation results.

TABLE V
ABLATION ANALYSIS OF GENERALIZED PANCREATIC CANCER SEGMENTATION ACROSS DIFFERENT MODULES.

| Method | RMYY | | | MSD | | | RENJI | | | Mean | | |
|---|---|---|---|---|---|---|---|---|---|---|---|---|
| | DSC (%) | ASD (mm) | HD (mm) | DSC (%) | ASD (mm) | HD (mm) | DSC (%) | ASD (mm) | HD (mm) | DSC (%) | ASD (mm) | HD (mm) |
| BL | 39.97 | 10.60 | 22.79 | 24.68 | 23.78 | 40.29 | 45.15 | 11.53 | 26.07 | 36.60 | 15.30 | 29.71 |
| BL+IDR | 44.12 | 10.43 | 20.98 | 27.16 | 26.37 | 40.14 | 41.30 | 12.07 | 27.70 | 37.52 | 16.29 | 29.60 |
| BL+DTL | 43.98 | 10.44 | 20.84 | 29.38 | 21.10 | 36.00 | 49.50 | 11.13 | 25.04 | 40.95 | 14.22 | 27.29 |
| BL+DSL | 46.32 | 10.19 | 20.03 | 32.37 | 19.92 | 33.20 | 51.87 | 10.84 | 24.24 | 43.52 | 13.65 | 25.82 |
| Ours | 48.36 | 8.62 | 17.67 | 36.85 | 18.06 | 28.60 | 53.12 | 8.98 | 22.10 | 46.11 | 11.88 | 22.79 |

BL: baseline, IDR: independent distance regression task, DL: dual-task learning, DSL: dual self-supervised learning.

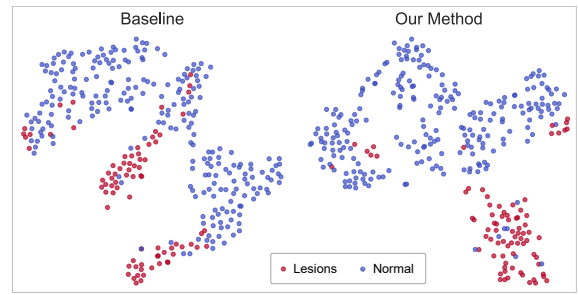

Fig. 6. t-SNE visualizations of the learned feature embeddings for the pancreatic lesions and the normal tissues.

TABLE VI
SEGMENTATION PERFORMANCE AND COMPUTATIONAL OVERHEAD COMPARISON OF OUR METHOD WITH OTHER STATE-OF-THE-ART GENERALIZATION METHODS IN CONSISTENT SETTINGS.

| Method | RMYY | | | MSD | | | RENJI | | | Mean | | | Computational Overhead | | |
|---|---|---|---|---|---|---|---|---|---|---|---|---|---|---|---|
| | DSC (%) | ASD (mm) | HD (mm) | DSC (%) | ASD (mm) | HD (mm) | DSC (%) | ASD (mm) | HD (mm) | DSC (%) | ASD (mm) | HD (mm) | Params. (×10$^7$) | GFLOPs* | Inf. Time (s/case)* |
| SNR | 39.95 | 12.48 | 24.97 | 22.23 | 28.07 | 36.27 | 41.81 | 12.29 | 27.14 | 34.66 | 17.61 | 29.45 | 3.79 | 946.02 | 11.59 |
| nnUNet_3d_lowres | 44.19 | 7.57 | 23.83 | 27.92 | 32.61 | 71.92 | 32.11 | 12.59 | 31.56 | 34.74 | 17.59 | 42.43 | 3.08 | 734.24 | 5.36 |
| DCAC | 36.48 | 11.76 | 24.35 | 25.63 | 25.62 | 41.23 | 44.78 | 11.68 | 26.55 | 35.63 | 16.35 | 30.70 | 3.79 | 945.98 | 11.65 |
| nnUNet_3d_fullres | 43.04 | 9.38 | 33.87 | 31.21 | 29.91 | 70.88 | 35.09 | 13.28 | 34.17 | 36.44 | 17.52 | 46.30 | 3.08 | 1003.68 | 22.90 |
| LANet | 39.02 | 12.66 | 25.66 | 28.20 | 22.74 | 37.89 | 42.26 | 12.81 | 27.93 | 36.49 | 16.07 | 30.49 | 3.79 | 958.14 | 11.42 |
| BigAug | 39.97 | 10.60 | 22.79 | 24.68 | 23.78 | 40.29 | 45.15 | 11.53 | 26.07 | 36.60 | 15.30 | 29.71 | 3.79 | 945.23 | 9.44 |
| PaNSegNet | 42.17 | 10.66 | 29.56 | 31.14 | 16.34 | 34.60 | 36.96 | 12.80 | 28.82 | 36.75 | 13.26 | 30.99 | 4.37 | 560.29 | 23.81 |
| HOG | 39.63 | 12.91 | 23.67 | 26.20 | 25.05 | 39.91 | 44.54 | 11.74 | 27.32 | 36.79 | 16.56 | 30.30 | 3.79 | 962.85 | 10.34 |
| nnUNet_3d_cascade | 44.37 | 9.15 | 31.40 | 31.66 | 29.29 | 73.41 | 35.56 | 12.61 | 31.81 | 37.19 | 17.01 | 45.54 | 12.30 | 1857.68 | 25.46 |
| Cutout | 40.62 | 12.53 | 24.19 | 26.34 | 24.80 | 39.98 | 47.81 | 11.27 | 25.42 | 38.25 | 16.20 | 29.86 | 3.79 | 945.23 | 10.22 |
| SAML | 41.85 | 11.48 | 22.65 | 27.62 | 24.17 | 39.23 | 46.26 | 11.89 | 26.59 | 38.57 | 15.84 | 29.49 | 3.79 | 945.23 | 9.90 |
| DAM | 41.60 | 10.20 | 21.35 | 28.43 | 22.95 | 38.21 | 46.62 | 9.82 | 23.75 | 38.88 | 14.32 | 27.77 | 3.79 | 945.12 | 9.96 |
| DoFE | 44.48 | 8.69 | 18.92 | 25.55 | 26.29 | 41.30 | 47.09 | 10.92 | 25.29 | 39.04 | 15.29 | 28.50 | 3.81 | 945.95 | 9.15 |
| MixStyle | 44.51 | 10.37 | 20.71 | 28.91 | 23.55 | 38.85 | 46.62 | 11.33 | 26.16 | 40.01 | 15.08 | 28.57 | 3.79 | 945.23 | 9.75 |
| DART | 47.14 | 9.39 | 18.36 | 30.84 | 27.25 | 38.04 | 44.30 | 11.38 | 26.29 | 40.76 | 16.00 | 27.56 | 7.58 | 1890.47 | 20.06 |
| RandConv | 46.07 | 10.27 | 20.22 | 34.28 | 20.51 | 32.01 | 47.30 | 10.90 | 25.05 | 42.55 | 13.89 | 25.76 | 3.79 | 945.41 | 9.98 |
| Ours | 48.36 | 8.62 | 17.67 | 36.85 | 18.06 | 28.60 | 53.12 | 8.98 | 22.10 | 46.11 | 11.88 | 22.79 | 3.80 | 945.44 | 10.52 |

GFLOPs*: We designate the inputs for coarse and fine segmentation as 64×224×224 and 32×128×128, respectively, except for PaNSegNet, which utilizes a fixed parameter configuration requiring an input size of 56×192×160.
Inf. Time (s/case)*: The time for fine segmentation, influenced by the ROI from coarse segmentation, may differ even with identical FLOPs.

### F. Ablation Analysis

We performed an ablation analysis on each component for the three datasets under consideration. The results, presented in Table V, demonstrate that the integration of an independent distance regression (IDR) task elevates the generalization capabilities of the baseline model (BL). Moreover, the dual-task synergistic learning (DTL) facilitates further enhancements in the model's generalization capacity. The dual self-supervised learning (DSL) module, by optimizing the stability of the model across both output and feature space, also advances the generalization performance. These findings underscore the pivotal role of each component in augmenting generalization capabilities. Additionally, the holistic model (Ours) that amalgamates all discussed components outperforms competing models, evidencing notable increments in the DSC by 8.39%, 12.17%, and 7.97%, respectively. In addition, as illustrated in Fig. 6, the t-distributed stochastic neighbor embedding (t-SNE) [46] visualization diagrams demonstrate that our method can effectively create stable enhanced representations of the tumor regions. These outcomes indicate that the proposed components are effective in enhancing generalized segmentation, and that incorporating all components leads to superior performance.

### G. Comparison with Other Methods

We present a comparative analysis of various state-of-the-art methods, namely, BigAug [17], Cutout [47], MixStyle [48], SAML [32], DoFE [34], SNR [49], DCAC [35], LANet [50], HOG [51], DAM [52], RandConv [38] and DART [53]. They were implemented in both the coarse and fine tumor segmentation phases, utilizing a uniform backbone network and pancreatic ROI. Besides, we incorporated the widely adopted nnUNet [54], which features three configurations: 3d_lowres, 3d_fullres, and 3d_cascade_fullres, along with one of its advanced variants PaNSegNet [45]. Table VI and Fig. 7 provide the quantitative results and the distance between tumor centroids in the annotations and those derived from each output, respectively. For nnUNet and PaNSegNet, their underlying principle is to construct appropriate data processing methods and architectures based on the specific characteristics of the data. This allows them to achieve excellent performance on data they have already seen. However, because these saettings are not designed for unseen data, the strategies used during training may not be well-suited for testing, leading to poor generalization performance. For multi-task learning methods [50-52], while they enhance robustness by concurrently addressing multiple objectives [55], their contributions to enhancing generalization capabilities often fall short when contrasted with methods expressly crafted for generalization purposes. Regarding the generalization techniques, although they have achieved remarkable success in certain tasks, their non-specific design for the task of pancreatic cancer pan-segmentation renders them ineffective in addressing three unique challenges associated with our task, thereby resulting in a generalization performance that is inferior to that of our proposed method. Moreover, Table VI and Fig.7 show that our method achieves superior performance in tumor localization and boundary prediction, with minimal errors in tumor centroids and the lowest ASD and HD. Thus, our method has a broad application potential in scenarios with arbitrary data availability, particularly with limited source domains.

## H. Computational Cost

Although modern hardware may mitigate computational costs in clinical workflows—such as the computationally intensive cascade 3D nnUNet processing about 1,000 scans in 8 hours with one GPU, outpacing many high-volume CT centers' daily throughput—computational efficiency remains critical for system scalability, deployment in resource-constrained settings (e.g., mobile or edge devices), and time-sensitive applications like intraoperative interventions or large-scale screening programs with high patient throughput. The computational cost (Table VI, encompassing both coarse and fine tumor segmentation), assessed by parameter count, inference time, and FLOPs, provides a measure of deployment demands. Unlike methods relying on a fixed backbone, nnUNet employs a data-adaptive model construction process, resulting in a distinct backbone. Besides, it employs a patch-based sliding window approach, processing each patch independently before reassembling them. This strategy inherently necessitates multiple inference passes per 3D volume, resulting in higher cost. Besides, computational efficiency varies minimally across other methods, except for DART. DART requires a separate network instance for each data center, leading to a parameter count proportional to the number of centers. Compared to the baseline, our method does not significantly increase parameters and maintains comparable inference time and FLOPs.

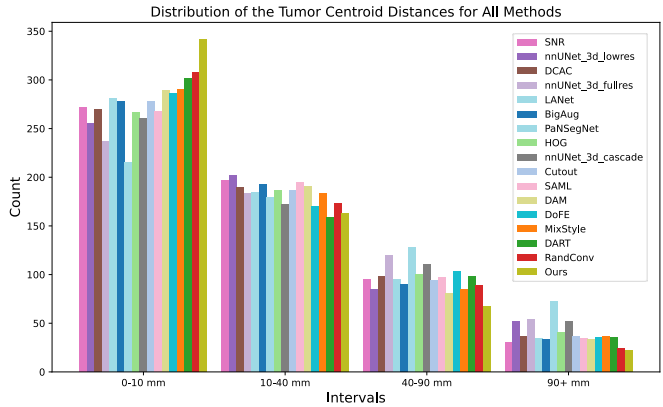

Fig. 7. The distribution of the tumor centroid distances between the segmentation results and the ground-truth labels.

## I. Failure Cases

The pancreatic cancer segmentation results in Table III reveal notably lower generalization performance on the MSD dataset compared to the other two datasets. This disparity among datasets can be attributed to the presence of biliary duct stents and different pancreatic lesions in the MSD dataset. In fact, this dataset contains common PDA samples, as well as IPMN and PNET samples. As reported in [56], 74 cases had biliary stents, while the remaining 207 samples included 65 IPMN and 36 PNET samples that differed significantly in their appearance from the PDA ones. Also, only 106 cases out of 281 cases in the MSD dataset are common PDA samples. This indicates the presence of disparity in pancreatic lesions that could greatly undermine the generalization performance. Fig. 8 further illustrates the challenge posed by this disparity, where cases with successful and poor segmentation outcomes have substantial appearance differences. The sample in the red box indicates that the model has recognized the tumor contours, but the difference in grayscale texture (due to different pathologies) prevented lesion identification by the model, led to an almost complete miss. In fact, generalization on IPMN and PNET samples can be considered as an extremely challenging cross-lesion generalization task. As a result, many cases in the MSD dataset could not be accurately segmented, and this resulted in significantly lower average results. These findings emphasize the need to address the challenge posed by the disparity among lesions in future work. In fact, we need to collect data for multiple pancreatic diseases and design models to mitigate interference from different pancreatic lesions.

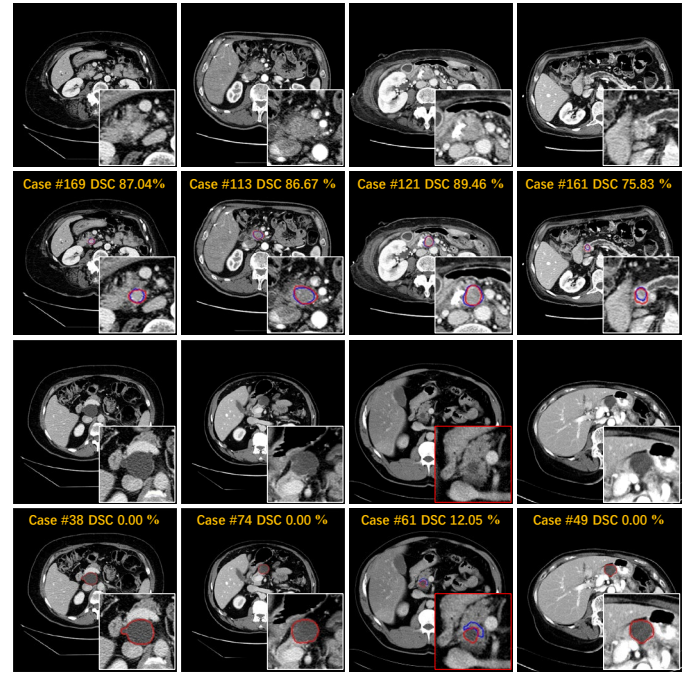

Fig. 8. Examples with excellent and poor generalized tumor segmentation in the MSD dataset. The red and blue contour denotes labels and segmentation results.

## V. DISCUSSION

The key to the treatment of pancreatic cancer, a highly malignant and lethal type of cancer, is early diagnosis and surgical resection. Automatic analysis methods provide important technical support towards this goal. However, factors such as variations in imaging characteristics and patient populations can lead to a significant performance degradation, limiting their applicability to unseen data. This is a grave limitation that hinders multi-center deployment in clinical practice and undermines the utility of model development efforts. Thus, this study focuses on developing a generalization framework that systematically addresses three specific generalization challenges in pancreatic cancer segmentation. Overall, the present work is intended to promote the deployment of enhanced segmentation models in wider clinical practice scenarios, supporting clinical needs such as early diagnosis, surgical planning, and personalized radiotherapy.

### A. Effectiveness of Mitigating the Three Main Challenges of This Task

The present study validates our method comprehensively and confirms its outstanding performance in mitigating the three aforementioned challenges. 1) For the challenge of significant appearance discrepancies, our approach remarkably enhances the adaptability and performance on the three datasets, as evidenced by the leading DSC values. Furthermore, Fig. 6 further highlights the improved stability of the tumor tissue representations in the feature space. 2) For the challenge of tumor contour variability, our method also demonstrates excellent performance in accurately fitting different tumor contours, as indicated by the lowest ASD and HD values obtained for both ablation and comparison experiments. 3) For the challenge of random anatomical locations, our method exhibits minimal location deviations, as depicted in Table VI. Also, the visual results in Fig. 7 highlight the improved adaptability to various anatomical locations. In summary, our approach effectively addresses these three major challenges, leading to considerable performance enhancements.

### B. Relationship Between Segmentation Performance and Hardware Specifications

In this study, each iteration was constrained to using only two new samples with the same anatomical structure but different grayscale textures (Fig. 2). This constraint was dictated by the hardware limitations of the number of GPUs and their memory size. In fact, this constraint can be adjusted to match the actual hardware device used, and hence optimize the segmentation performance. The generation of more samples with the same structure but with different grayscale textures in each iteration can help model to learn to handle a wider range of appearance variations simultaneously. This potentially enhances representation stability and segmentation quality. Notably, the number of generated samples in each iteration does not affect the computational efficiency during the inference phase. Hence, it is feasible to train the model on a better computing platform and then deploy it on common devices. In practical applications, we can balance the model performance and training efficiency through the utilization of reasonable resource allocation.

### C. Comparison with distance maps-based methods

We have observed that there exist some studies [57-59] on segmentation algorithms based on distance regression, such as employing distance maps to construct a pancreatic skeleton to enhance segmentation stability [58], and utilizing distance regression to provide additional supervision to mitigate semi-supervised challenges [59]. While these studies have achieved remarkable performances, they have primarily focused on the role of distance regression within the realms of conventional intra-domain segmentation or semi-supervised segmentation tasks, overlooking its potential in augmenting the generalizability. Moreover, these approaches often segregate the segmentation and regression tasks, without attempting to establish a connection between them, thus not fully exploiting the potential of distance regression tasks. In our work, we introduce a dual-task synergy framework that successfully facilitates collaboration between these two tasks. Compared to merely constructing independent regression tasks, our approach significantly enhances the model's generalization stability across various datasets (Table VI).

### D. Methodological Limitations and Future Work

This study jointly handles pancreas segmentation and pancreatic cancer segmentation. The pancreas is segmented to define a pancreatic ROI, followed by tumor segmentation within this ROI. Although this scheme is commonly used for small target segmentation [60], executing multiple models sequentially may impact real-time performance. The widely used 3D patch matching method may increase the inference times due to the invocation of a sliding window test across three spatial dimensions. Thus, the exploration of methods that can segment both the pancreas and the pancreatic lesion is definitely encouraged to attain real-time performance.

Due to the challenging nature of pancreatic cancer segmentation, there may be substantial differences in manual annotation among different clinicians. In fact, clinicians may differ in their preferences for conservative or radical tumor contours. This variation in the labeling criteria can impact the accuracy of generalized segmentation results. Two experienced clinicians [56] separately annotated several cases from the MSD dataset, achieving a DSC of only 63%. This highlights the impact of subjective judgment on tumor annotation. To mitigate the undesirable effects of annotation preferences for different datasets, this study employed different thresholds: 0.9 for the RMYY dataset, 0.7 for the MSD dataset, and 0.1 for the RENJI dataset. In clinical practice, cases are often annotated by two clinicians, with disagreements resolved by a senior clinician. This practice increases workload and can be impractical for many datasets. Thus, one of the future research directions will focus on leveraging large datasets and advanced models to aid in the annotation process, thereby mitigating the impact of label variability. This approach aims to establish a unified standard for cross-center generalization evaluation, ultimately enhancing the reliability and credibility of related research findings.

This study used venous-phase CT data from three different centers; however, the sample size remains relatively small. This limitation could potentially constrain our ability to draw broad conclusions. Given that enhanced CT images are typically divided into four phases—non-contrast, arterial, venous, and delayed—the varying timing of contrast agent administration in each phase can significantly influence imaging styles. To address this, incorporating data from other phases and imaging modalities, such as arterial-phase CT images and MRI with T1- and T2-weighted sequences, is expected to contribute to enhancing generalization performance. Thus, expanding dataset to include more cases and a broader range of imaging modalities will be crucial for further validating and enhancing the generalizability of our approach across diverse clinical settings. Additionally, we plan to develop scalable generalization strategies that are tailored to the dataset's composition and size. This will enable optimal utilization of available data at different stages and offer potential solutions for medical institutions with limited data resources.

## VI. Conclusion

In the realm of pancreatic cancer segmentation, prevailing methods often grapple with the challenge of deteriorated generalization performance under real-world scenarios. This work addresses three primary challenges of the generalization problem through our dual-task synergy-driven generalization framework. The former is used to effectively address the variations in tumor contours and their spatial distributions, while the latter enhances the stability in representation and segmentation across varying grayscale textures in both the feature and output spaces. Our method effectively improves the generalization performance of the entire framework and enhances its potential for a wider range of practical scenarios. Moreover, our method is model-independent and can be easily integrated into other similar tasks. Hence, the proposed method has the potential to serve as a potent computational tool for expeditious disease detection, high-precision diagnosis, and personalized treatment, which is expected to advance the automation of the associated medical procedures.